\documentclass[namedreferences]{solarphysics}
\usepackage[natbib,optionalrh,solaromanenum]{spr-sola-addons} 
\usepackage{graphicx}        
\usepackage{amssymb}        
\usepackage{color}           
\usepackage{url}             
\usepackage{multirow}

\newcommand{\aap}{    {\it Astron. Astrophys.}}

\newcommand{\apj}{    {\it Astrophys. J.}}
\newcommand{\apjl}{   {\it Astrophys. J. Lett.}}

\newcommand{\solphys}{{\it Solar Phys.}}

\newcommand{\raa}{    {\it Res. Astron. Astrophys.}}
\newcommand{\apjs}{    {\it Astrophys. J. Suppl. Ser.}}

\begin{document}

\begin{article}

\begin{opening}

%
\title{Investigation of Umbral Dots with the \textit{New Vacuum Solar Telescope}}

%
\author{Kaifan ~\surname{Ji}$^{1}$\sep
        Xia ~\surname{Jiang}$^{1,2}$\sep
        Song ~\surname{Feng$^{1,2}$}\sep
        Yunfei ~\surname{Yang}$^{1,2}$\sep
        Hui ~\surname{Deng}$^{1}$\sep
        Feng ~\surname{Wang$^{1}$}\sep
       }

%
\runningauthor{Ji et al.}
\runningtitle{Investigation of Umbral Dots with the \textit{New Vacuum Solar Telescope}}

%
  \institute{$^{1}$ Faculty of Information Engineering and Automation/Yunnan Key Laboratory of Computer Technology Application, Kunming University of Science and Technology, Kunming 650500, China
      email: \url{ynkmfs@escience.cn}\\
   $^{2}$ Key Laboratory of Solar Activity, National Astronomical Observatories, Chinese Academy of Sciences, Beijing 100012,China\\
             }

\begin{abstract}

Umbral dots (UDs) are small isolated brightenings observed in sunspot umbrae. They are convective phenomena existing inside umbrae. UDs are usually divided into central UDs (CUDs) and peripheral UDs (PUDs) according to their positions inside an umbra. Our purpose is to investigate UD properties and analyze their relationships, and further to find whether or not the properties depend on umbral magnetic field strengths. Thus, we selected high-resolution TiO images of four active regions (ARs) taken under the best seeing conditions with the \textit{New Vacuum Solar Telescope} in the Fuxian Solar Observatory of the Yunnan Astronomical Observatory, China. The four ARs (NOAA 11598, 11801, 12158, and 12178) include six sunspots. A total of 1220 CUDs and 603 PUDs were identified. Meanwhile, the radial component of the vector magnetic field of the sunspots taken with the \textit{Helioseismic and Magnetic Imager} on-board the \textit{Solar Dynamics Observatory} was used to analyze relationships between UD properties and umbral magnetic field strengths. We find that diameters and lifetimes of UDs exhibit an increasing trend with the brightness, but velocities do not. Moreover, diameters, intensities, lifetimes and velocities depend on the surrounding magnetic field. A CUD diameter was found larger, the CUD brighter, its lifetime longer, and its motion slower in a weak umbral magnetic field environment than in a strong one.
\end{abstract}
\end{opening}

\section{Introduction}
There are some small bright features called umbral dots (UDs) in a dark umbra, and the features can be found almost all over umbrae and pores. UDs only cover 3 -- 10 \% of the umbral area, but contribute 10 -- 20 \% of the brightness. Therefore, convective motions must exist within umbrae because radiation cannot explain the phenomena \citep{1965ApJ...141..548D,1993ApJ...415..832S}. Therefore, UDs play a vital role in the energy balance of sunspots. The study of UDs is crucial to understand convective motions and interactions of the plasma with strong magnetic fields, and analyze the formation mechanism of sunspots. Usually, UDs are divided into two classes according to their origins \citep{1986A&A...156..347G}. UDs are called peripheral UDs (PUDs) near an umbra--penumbra boundary and central UDs (CUDs) in an umbral center. PUDs are generally brighter than CUDs, and move quickly toward the umbral center, while CUDs are relatively static. Two different models have been proposed to explain the formation mechanism of UDs: the clustered magnetic flux tube and the monolithic flux tube model. The former considers that UDs represent hot field-free gas intruded into a cluster of magnetic flux tubes \citep{1979ApJ...234..333P}. The latter suggests that the energy transport in an umbra is dominated by non-stationary narrow rising plumes of hot plasma with adjacent down-flows. UDs are formed by the narrow up-flow plumes that become almost field-free near the surface layer \citep{2006ApJ...641L..73S}. The essential difference between the two models is along the boundary of the UDs there exist local down-flows in the monolithic flux tube model. Detailed studies of UD properties, such as morphologies, velocities, lifetimes, and intensities, and relationships between different properties are crucial to understand the nature of the local convective motions \citep{1997A&A...328..682S,1997A&A...328..689S,2007ApJ...669L..57B, 2008A&A...492..233R,2012ApJ...757...49W}.

Our purpose is to analyze UD properties and relationships between different properties, and further to find whether or not the properties depend on umbral magnetic field strengths. So we selected high-resolution observations of four active regions (ARs) taken under the best seeing conditions with the \textit{New Vacuum Solar Telescope} (NVST) in the Fuxian Solar Observatory of the Yunnan Astronomical Observatory, China.

The organization of the paper is as follows. The observations and data reduction are described in Section 2. Section 3 briefly describes the identification procedure of CUDs and PUDs. Section 4 illustrates UD properties and their relationships. Moreover, the relationships between UD properties and umbral magnetic field strengths are discussed. Finally, we give our conclusions in Section 5.

\section{Observations and Data Reduction}
The NVST is a vacuum solar telescope with a 985-mm clear aperture whose purposes are to obtain high-resolution imaging and spectral observations, including measurements of the solar magnetic field. The telescope consists of one channel for the chromosphere and two channels for the photosphere. The band used for observing the chromosphere is H$\alpha$ (656.3$\pm$0.025 nm). The bands for observing the photosphere are TiO (705.8$\pm$1 nm) and G-band (430.0$\pm$0.8 nm). The high-resolution data of the NVST are classified into two levels. The level 1 data are processed by frame selection (lucky imaging) \citep{2004Obs...124..159T}. The level 1+ data are reconstructed by speckle masking \citep{1983ApOpt..22.4028L} or iterative shift and add \citep{1998SPIE.3561.....Z}. Technical details of the NVST were described by \citet{2014RAA....14..705L} and \citet{2014IAUS..300..117X}.

\begin{figure}
        \centering
        \includegraphics[width=8cm]{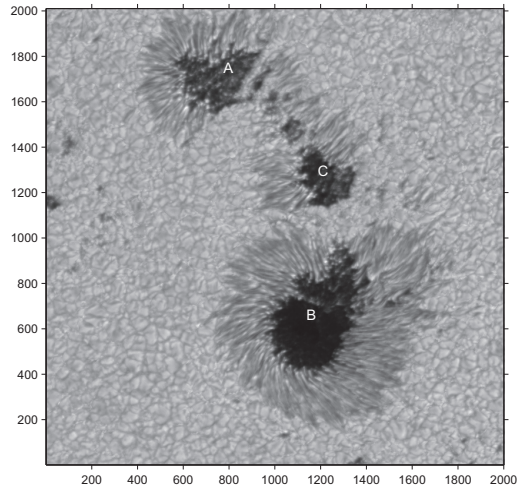}
        \caption{A reconstructed and projection-corrected image of NOAA 12158 recorded on 13 September 2014 at 03:00:00 UT with the NVST. The active region included three sunspots marked as A, B, and C.}
        \label{fig1}
\end{figure}

Because the TiO line is highly sensitive to umbral temperature variations \citep{2003A&A...412..513B}, its observations are more appropriate to investigate UDs. Therefore, we used the level 1+ TiO observations, rather than the H$\alpha$ and G-band ones. We selected high-resolution image sequences of six sunspots taken under the best seeing conditions since October 2012. The sunspots located in four ARs: NOAA 11598, 11801, 12158, and 12178. The observation parameters are listed in Table 1. The high-resolution data were obtained without adaptive optics. But the resolution of the reconstructed images can almost reach as high as the diffraction limit of the NVST in good seeing conditions \citep{2014RAA....14..705L}. From the table we see that the pixel size of the sequences is different before and after 2014. This is because the NVST team changed their optical system on 19 May 2014.

The images in each sequence were co-aligned by the subpixel registration algorithm \citep{feng2012subpixel, 2015RAA....15..569Y}. From Table 1 we see that the sunspots located in NOAA 11598, 11801, and 12158 were away from the solar disk center, especially the sunspot of NOAA 11801. Its heliocentric angle, $\theta$, was close to 39$^\circ$ (\textit{i.e.}, cos($\theta$)=0.78). So we used the SolarSoft routine \textit{wcs\_convert\_from\_coord} to transform the images of NOAA 11598, 11801, and 12158 to heliographic coordinates for correcting the projection effects \citep{2013arXiv1309.2392S}. Figure 1 shows a reconstructed and projection-corrected image of NOAA 12158 recorded on 13 September 2014 at 03:00:00 UT, where three sunspots (A, B, and C) are present.
\begin{table}
\caption{Parameters of the observed sunspots}
\begin{tabular}{cccccc}
\hline
AR NOAA  &  Date (d-m-y)      &  Time interval (UT)        & Location & Pixel size (") & Cadence(s)\\
\hline
11598  	 & 29-10-2012   &05:50:13--07:44:36   & S11W27    &   0.041      &37 \\
\hline
11801  	 & 01-08-2013   &03:38:35--04:34:55   & W31N24    &    0.041     &20\\
\hline
12158  	 & 13-09-2014   &02:57:30--03:47:14   & N20W27    &   0.052      &30\\
\hline
12178  	 & 03-10-2014   &04:35:00--05:33:03   & S01E05    &   0.052       &40\\
\hline
\label{tbl1}
\end{tabular}
\end{table}

\section{Identification and Tracking of UDs}
We followed the method discussed in \citet{2015SoPh..290.1119F} to identify and track CUDs and PUDs. The method mainly consists of three steps: first, the periphery (umbra--penumbra) and center boundaries in an umbra are detected based on the morphological reconstruction technique; second, the UDs are identified based on the phase congruency technique; finally, the identified UDs are tracked based on a 26-connected neighborhood technique. The phase congruency technique has been used to extract low-contrast solar features, like coronal loops and umbral flashes \citep{2014SoPh..289.3985F,2014RAA....14.1001F}.

Empirical distance thresholds have been used to divide UDs into CUDs and PUDs \citep{2008A&A...492..233R, 2009ApJ...702.1048W, 2011SoPh..270...75H, 2012ApJ...752..109L}. \citet{2008A&A...492..233R} considered an UD as a PUD if the UD's birth position is closer than 400 km from a defined umbra boundary, otherwise it is defined as a CUD. \citet{2011SoPh..270...75H} defined a narrow width near an umbral boundary where the UDs are considered as PUDs. \citet{2012ApJ...752..109L} considered UDs with a threshold 0.8" inward from an assigned umbra boundary as CUDs, and those located outward as PUDs.

Because the definition of the periphery (umbra--penumbra) and center boundaries is crucial to classify UDs, we briefly introduce the definition and the identification procedure proposed by \citet{2015SoPh..290.1119F}. They defined the periphery and center boundaries according to an umbral profile. We show the three-dimensional surface of a reconstructed and projection-corrected image obtained from NOAA 11801 in Figure 2. The X and Y axes indicate the image size and the Z axis its normalized intensity. The color bar on the right indicates the intensity range from 0 (blue) to 1 (red). As illustrated with a red dashed line, the profile of the umbra can be approximated by a trapezoid whose two sides appear skewed, and the base is relatively flat. Here, the base of the trapezoid is defined as the center of the umbra where the UDs are considered as CUDs. The two skewed sides of the trapezoid are defined as the peripheral region where the UDs are considered as PUDs. For obtaining the periphery and center boundaries, an image was reconstructed by the morphological reconstruction technique, and two thresholds, 0.3 and 0.6 $R_{max}$, were used. $R_{max}$ denotes the maximum intensity of the reconstructed image.

Identified UDs were divided into CUDs and PUDs according to the boundaries. However, we only obtained an insignificant number of PUDs in sunspots A, B, and C of NOAA 12158. The PUDs are so few that we cannot obtain accurate statistical results. Therefore, we abandoned the PUDs in these three sunspots. As a result, we extracted CUDs from six sunspots and PUDs from three sunspots. Although the division method of CUDs and PUDs might mis-classify a few CUDs or PUDs, this has no effect on statistical results of UDs. Two identified results are shown in Figure 3. The periphery and center boundaries are marked with yellow and red curves, and the UDs are marked with white contours.

\begin{figure}
        \centering
        \includegraphics[width=8cm]{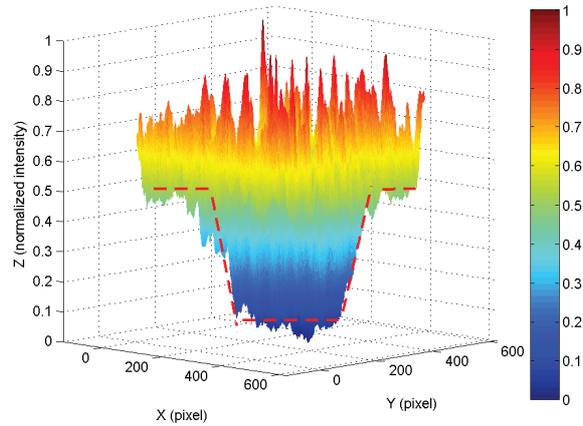}
        \caption{The 3D surface of a reconstructed and projection--corrected spot in AR 11801 whose original image was recorded on 1 August 2013 at 03:48:04 UT. The X and Y axes indicate the image size and the Z axis its normalized intensity. The color bar on the right indicates the intensity range from 0 (blue) to 1 (red). The red dashed line illustrates the umbral profile that can be approximated by a trapezoid.}
        \label{fig2}
\end{figure}

\begin{figure}[t]
        \centering
        \includegraphics[height=4.5cm]{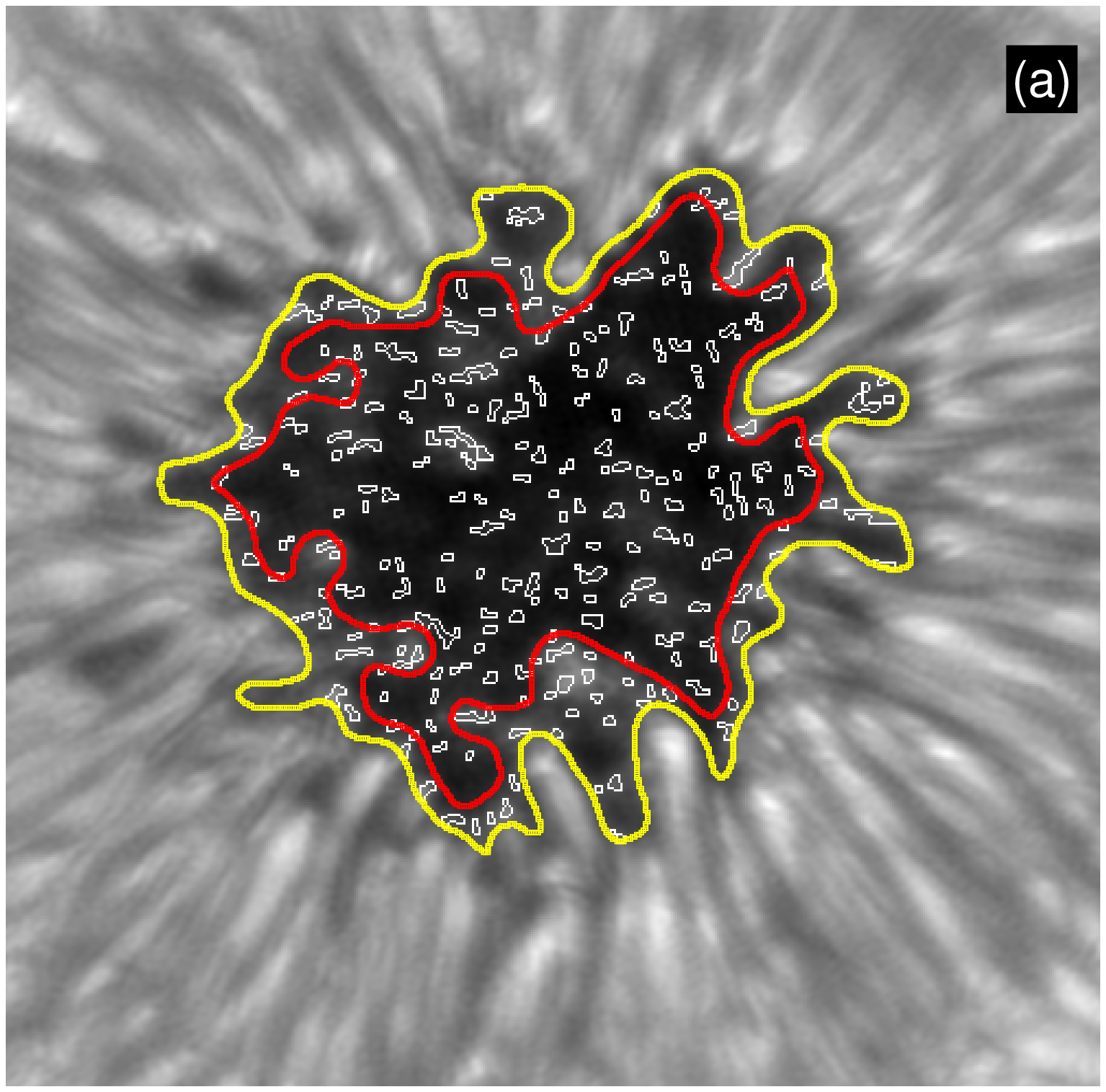}
        \includegraphics[height=4.5cm]{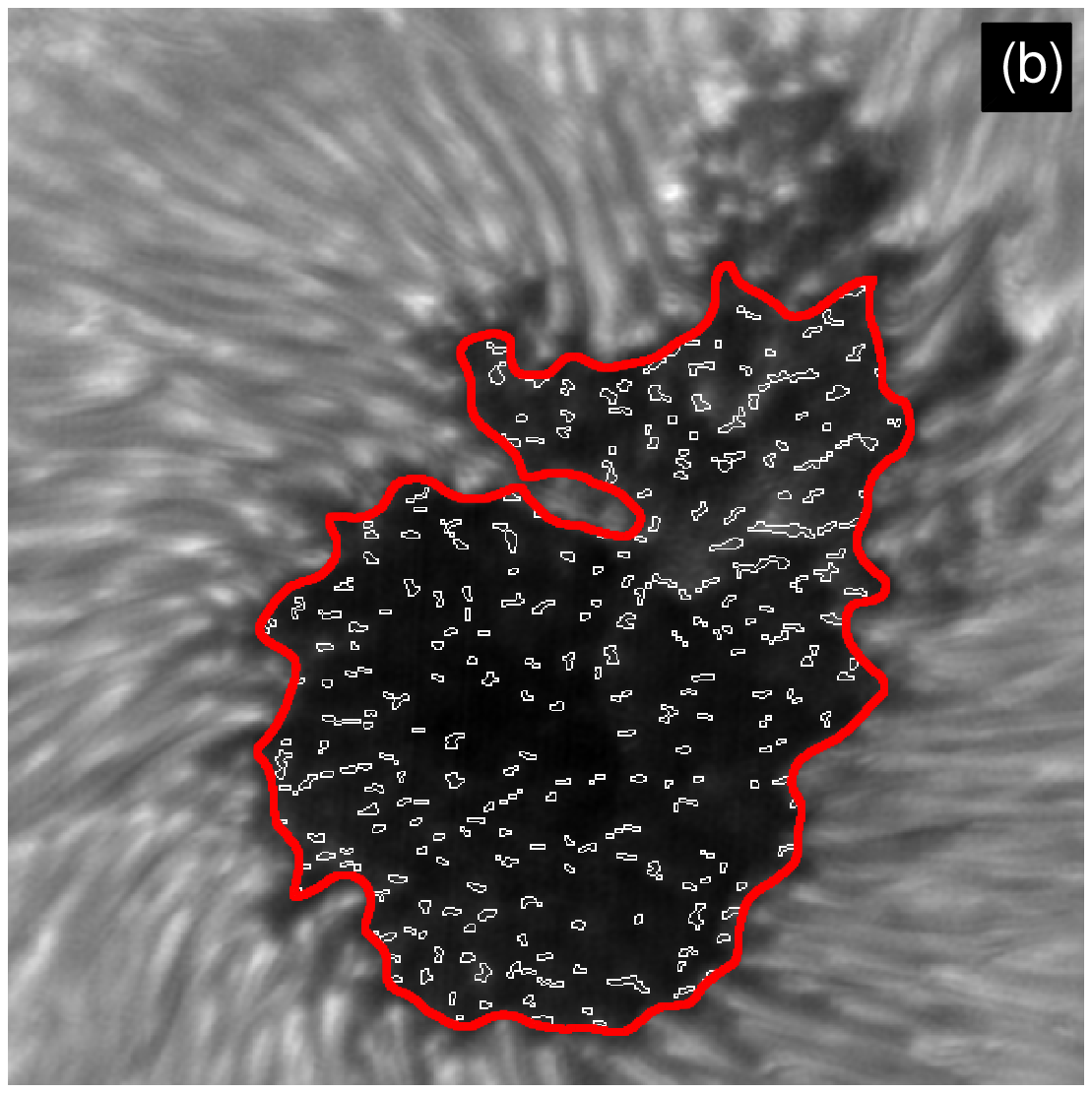}
        \caption{Boundaries identified by the method \citep{2015SoPh..290.1119F}. The periphery (umbra--penumbra) and center boundaries are marked with yellow and red curves, and the UDs are marked with white contours. (a) The result for a corrected sample image from NOAA 11801 whose original image was taken on 1 August 2013 at 03:48:04 UT. (b) The result for the sunspot marked as B in Figure 1. Note that there is no yellow curve. This is because only an insignificant number of PUDs were identified, and they are not enough to obtain accurate statistical results. Thus, we abandoned the PUDs and the corresponding peripheral region.}
        \label{fig3}
\end{figure}

\section{Results and Discussion}
Firstly, the means and standard deviations of UD properties such as equivalent diameters, ratios of the maximum intensity to the average intensity of the corresponding adjacent umbral background, lifetimes, and horizontal velocities are obtained with distribution functions. Secondly, the intensity--diameter, lifetime--diameter, lifetime--intensity, and velocity--intensity relationships are analyzed. Finally, the relationships between UD properties and umbral magnetic field strengths are discussed.

\subsection{Property Definition and Feature Extraction}

For each UD, its equivalent diameter $(D_{eq})$ is calculated as $\sqrt{4A/\pi}$, where $A$ denotes the total number of pixels of an UD. The ratio of the maximum intensity $(I_{ud})$ to the average intensity of the adjacent umbral background $(I_{bg})$ are determined from the identified areas. Some UD properties are defined using the tracking procedure; they are lifetime $(T_{ud})$, birth--death distance $(L_{bd})$, and horizontal velocity $(V_{ud})$. $T_{ud}$ is the sum of the cadence of all frames; $L_{bd}$ is the centroid distance from its birth to death frame; $V_{ud}$ is computed by dividing $L_{bd}$ by $T_{ud}$.

We rejected the UDs whose diameters were lower than 130 km (0.18") and lifetimes were less than two minutes for accurate statistical results. In the tracking procedure, if splitting or merging occurred, the UD was discarded. As a result, a total of 1220 CUDs and 603 PUDs were identified. Table 2 lists the number of CUDs and PUDs of each sunspot. Subsequently, we utilize them to analyze UD properties and their relationships, and the relationships between these properties and umbral magnetic field strengths.

\begin{table}
\caption{The number of the identified UDs of each sunspot}
\begin{tabular}{ccccccc}
\hline
AR NOAA     & 11598     &11801      &\multicolumn{3}{c}{12158}      &12178   \\
\hline
Spot &  &  & A&B&C &  \\
\hline
CUD &313 	    & 201      & 90&279&75&262 \\
\hline
PUD &219 	    & 162      &  &   &   &222 \\
\hline
\label{tbl2}
 \end{tabular}
\end{table}

\subsection{UD Properties}

We obtained the probability histograms of UD properties in each sunspot, and fitted them with distribution functions for obtaining property values of UDs. Figure 4 shows the histograms and their fit curves of the UDs located in AR NOAA 12178; other ARs are not shown due to similar histograms and fit curves. But all the fitted means and standard deviations are listed in Table 3. The red color indicates the histograms and the corresponding fit curves of the CUDs, and the blue color those of the PUDs in Figure 4.

\begin{table}
\caption{All means and standard deviations of UD properties}
\begin{tabular}{cccccc}
\hline
\multicolumn{6}{c}{CUDs}\\
\hline
NOAA        &Spot    & Diameter(km)          & $I_{ud} / I_{bg}$      &Lifetime(min)           & Velocity(km $s^{-1}$)  \\
\hline
11598  	    &   & 178$\pm$40     &1.05$\pm$0.02        &5.35        &0.37$\pm$0.19 \\
\hline
11801  	    &   & 216$\pm$38     &1.07$\pm$0.03        &4.48        &0.30$\pm$0.16 \\
\hline
12158  	    &A   & 235$\pm$41     &1.10$\pm$0.05        &5.67       &0.27$\pm$0.14 \\
\hline
12158  	    &B   & 225$\pm$40     &1.08$\pm$0.04        &5.66       &0.38$\pm$0.20\\
\hline
12158  	    &C   & 234$\pm$45     &1.09$\pm$0.04        &7.26       &0.19$\pm$0.10\\
\hline
12178  	    &   & 210$\pm$38     &1.07$\pm$0.03        &4.59        &0.37$\pm$0.19\\
\hline
\multicolumn{6}{c}{PUDs}\\
\hline
11598  	    &   & 195$\pm$41     &1.08$\pm$0.06         &6.25    &0.47$\pm$0.24 \\
\hline
11801  	    &   & 226$\pm$41     &1.12$\pm$0.05         &8.12   &0.51$\pm$0.27\\
\hline
12178  	    &   & 226$\pm$46     &1.15$\pm$0.09         &7.95    &0.45$\pm$0.24 \\
\hline
\label{tbl3}
\end{tabular}
\end{table}

\begin{figure}
        \centering
        \includegraphics[width=6cm]{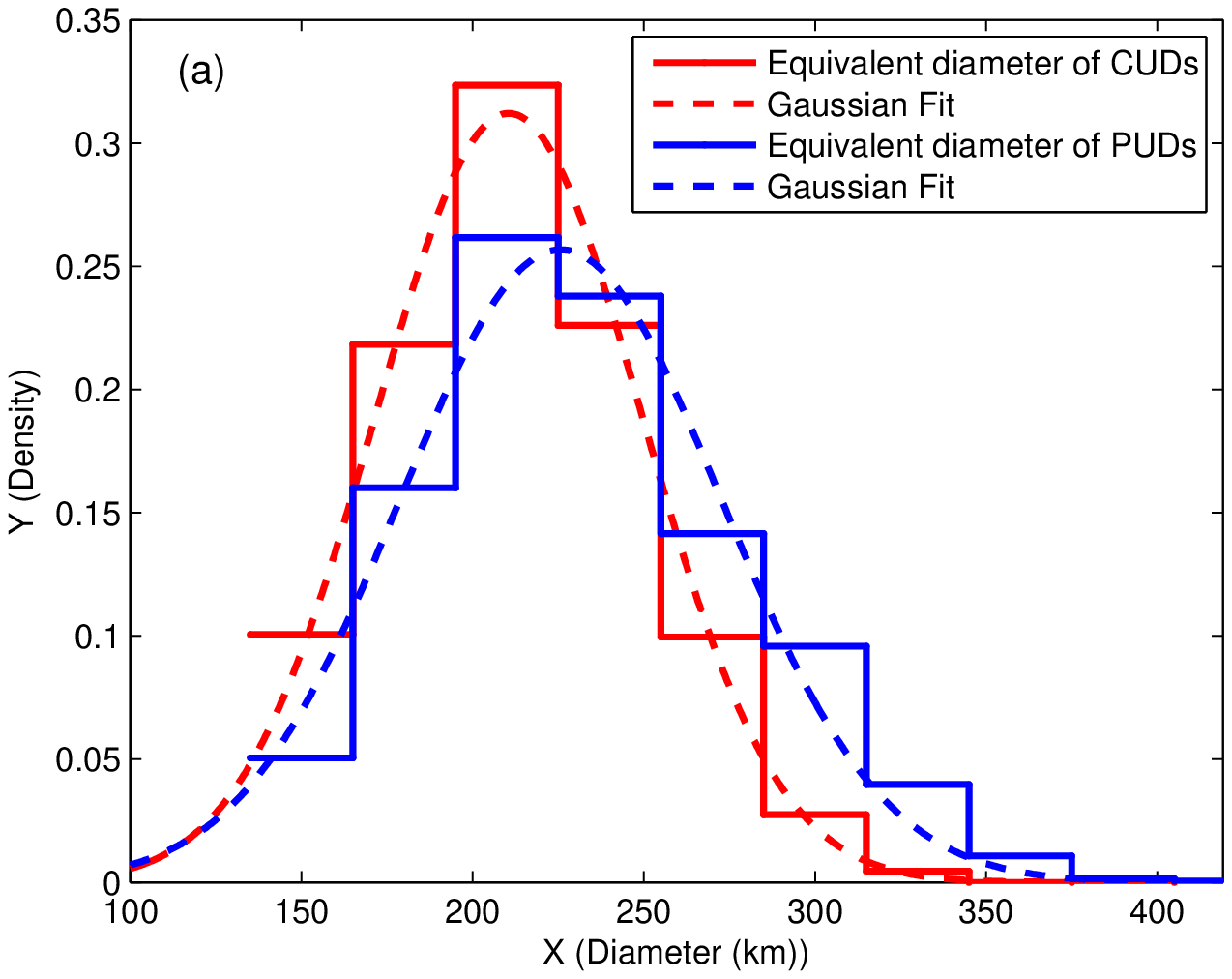}
        \includegraphics[width=6cm]{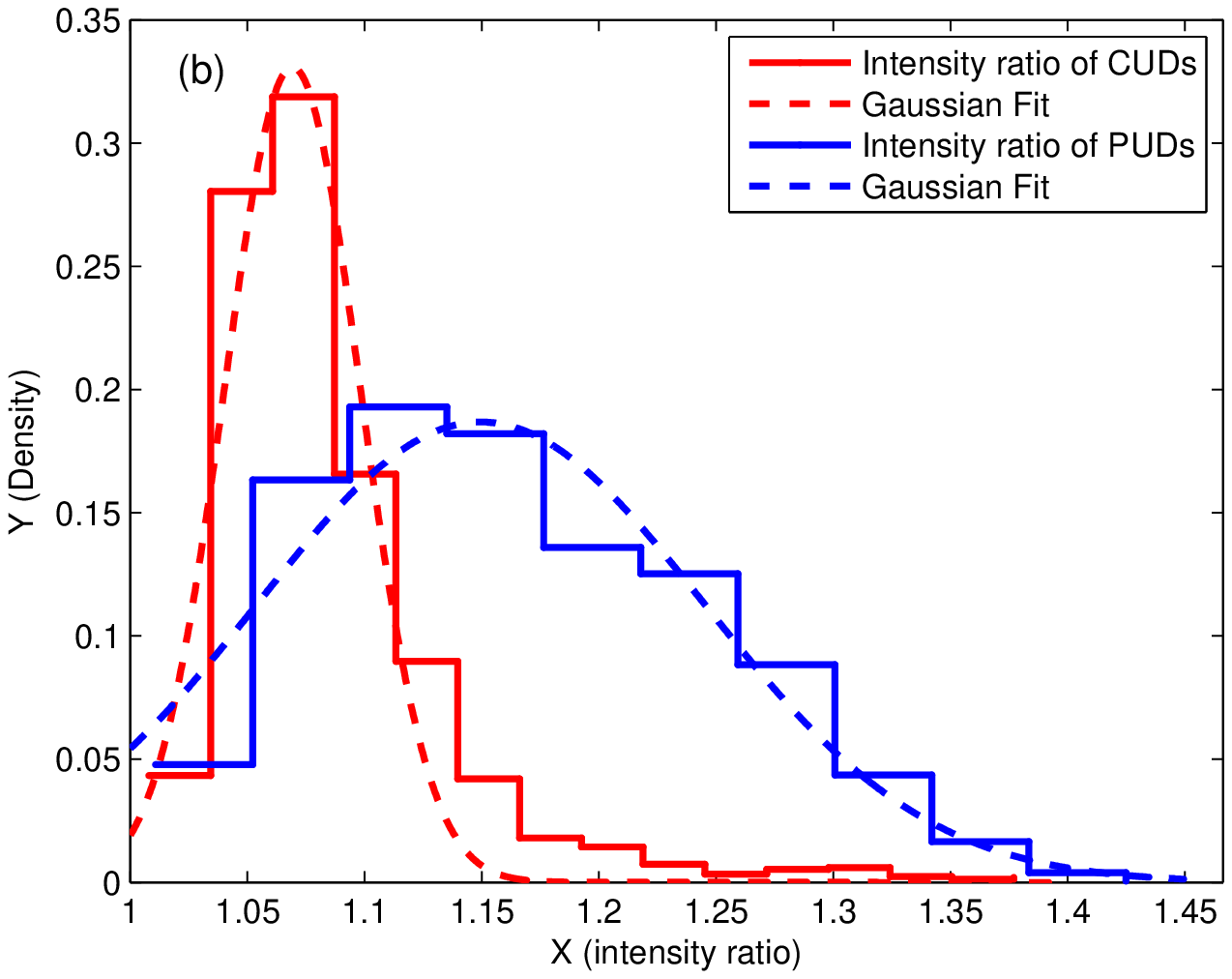}
        \includegraphics[width=6cm]{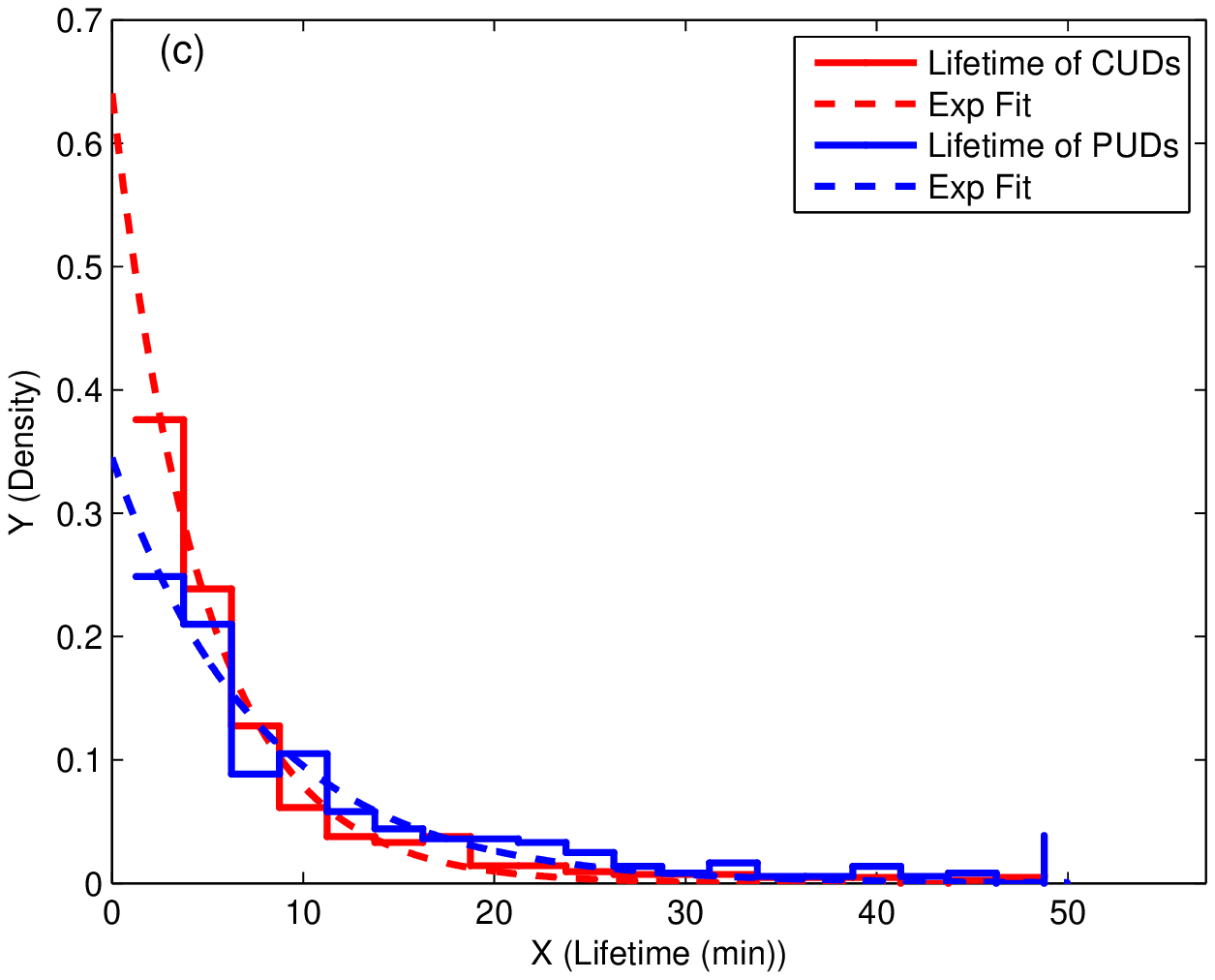}
        \includegraphics[width=6cm]{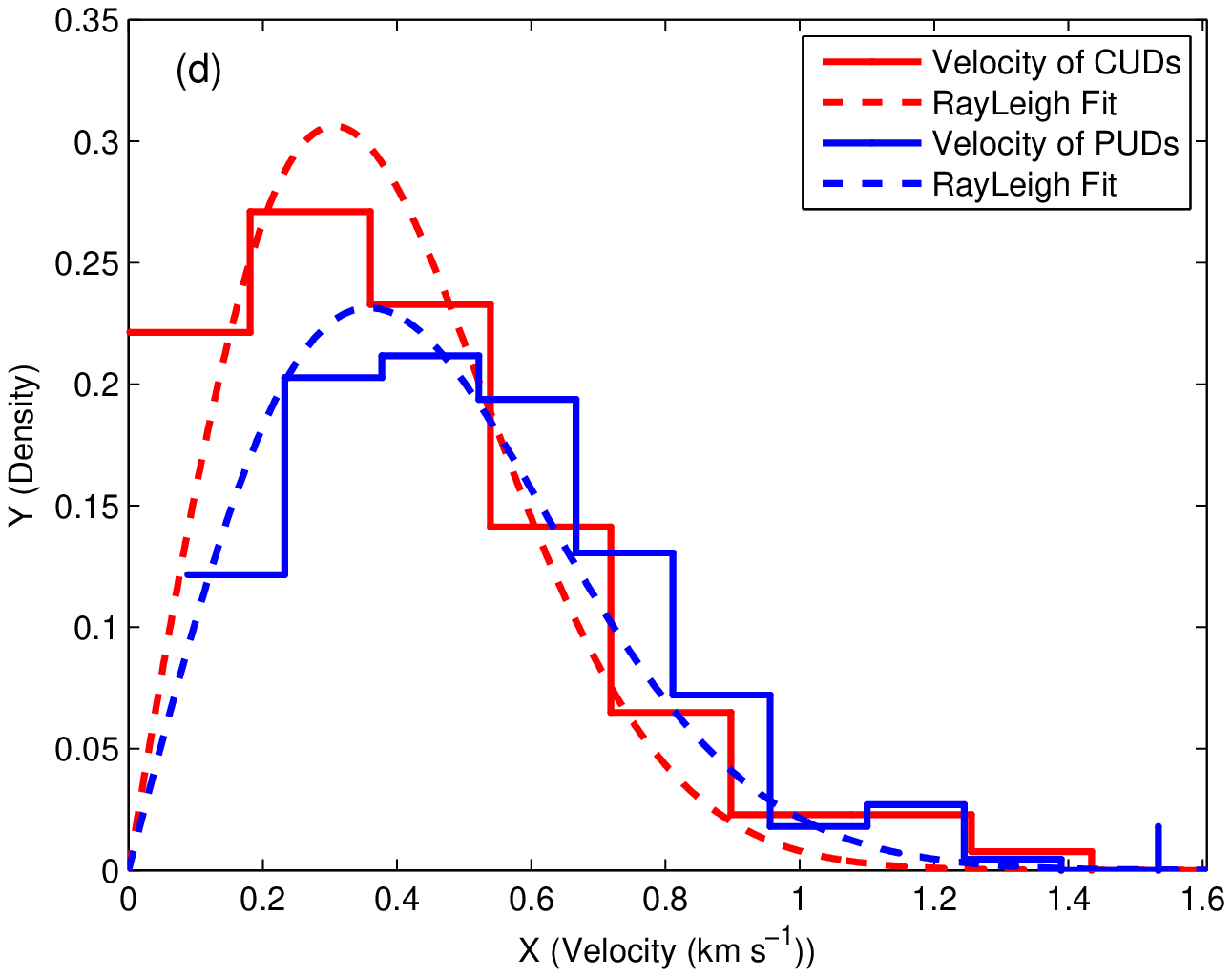}
        \caption{Probability histograms and fit curves of the UDs located in the sunspot of NOAA 12178. (a) Equivalent diameter, (b) intensity ratio, (c) lifetime, and (d) horizontal velocity. The red color indicates the histograms and fit curves of CUDs, and the blue color those of PUDs. }
        \label{fig4}
\end{figure}

Figure 4a and b show the probability histograms of equivalent diameters and intensity ratios. We followed the method used in previous works \citep{2011SoPh..270...75H,2012ApJ...757...49W,2015SoPh..290.1119F} to fit the distribution of the diameters and the intensity ratios with a Gaussian function. The function is as follows.
\begin{equation}\label{Eq. 1}
    y=\frac{1}{\sigma\sqrt{2\pi}}e^{-\frac{(x-\mu)^2}{2\sigma^2}}.
\end{equation}

Here, $\mu$ and $\sigma$ are the mean and the standard deviation, respectively. $\mu$ was used as the property values of equivalent diameters and intensity ratios. The mean diameters of CUDs range from 178 to 235 km, and that of PUDs from 195 to 226 km (see the third column in Table 3). The diameters of all the UDs range from 131 to 366 km. The results are in good agreement with those of previous works \citep{2008SoPh..250...17H, 2008A&A...492..233R, 2010A&A...510A..12B, 2012ApJ...745..163K}. They concluded that the range of UD diameters is between 150 and 350 km. The mean intensity ratios of CUDs range from 1.05 to 1.10 and that of PUDs from 1.08 to 1.15 (see the fourth column in Table 3). This demonstrates that the brightness of CUDs is 5 -- 10 \% higher than that of their adjacent backgrounds, while the brightness excess is 8 -- 15 \% for PUDs.

As shown in Figure 4c, the probability histograms of UD lifetimes approximate an exponential distribution. \citet{2008A&A...492..233R, 2011SoPh..270...75H, 2015SoPh..290.1119F} also found that UD lifetimes exhibit a similar distribution. So we used an exponential function to fit the histograms, which is described as follows.

\begin{equation}\label{Eq. 2}
    y= \lambda e^{-\lambda x}.
\end{equation}

For an exponential distribution, both the mean and the standard deviation are equal to $\lambda^{-1}$, which was used as the property value of UD lifetimes. The mean lifetimes of UDs range from 4 to 8 minutes (see the fifth column in Table 3). This result is in qualitative agreement with other works in the literature \citep{1999ApJ...511..436S, 2008A&A...492..233R, 2011SoPh..270...75H, 2015SoPh..290.1119F}.

The horizontal velocity histograms of UDs, regardless of CUDs and PUDs, show approximately Gaussian distribution with zero mean and the same variance in the X and Y axes of the helioprojective--Cartesian coordinate system, demonstrating that the horizontal velocity follows a Rayleigh distribution \citep{siddiqui1964statistical}. Therefore, we used a Rayleigh function to fit the velocity histograms. The function is defined as,
\begin{equation}\label{Eq. 2}
    y= \frac{x}{\sigma^2} e^{-\frac{x^2}{2\sigma^2}},
\end{equation}
where $\sigma$ is the scale parameter of the distribution. The mean and the standard deviation are equal to $\sigma\sqrt{\pi /2}$ and $\sigma\sqrt{(4-\pi)/2}$, respectively. We used the two values to describe the velocity property of UDs, rather than the peak of the distribution. The mean values and standard deviations are listed in column 6 of Table 3. The velocity range of CUDs is between 0.19 and 0.38 km $s^{-1}$ and that of PUDs between 0.45 and 0.51 km $s^{-1}$ (see Table 3); the horizontal velocity of CUDs is slower than that of PUDs. \citet{2009ApJ...702.1048W} and \citet{2012ApJ...745..163K} concluded that PUDs have a relatively fast horizontal velocity because they are located in regions with a strong horizontal component of the magnetic field and/or a strongly inclined field. Our findings confirm their conclusions.

\subsection{Correlation Analysis of UD Properties}
The relationships, such as intensity--diameter, lifetime--diameter, lifetime--intensity, and velocity--intensity, for all CUDs (1220, obtained from six sunspots) and all PUDs (603, from three sunspots) are presented in the form of scatter plots in Figure 5.

For obtaining the trend of the scatter points, the points were sorted into 15 equally spaced bins along the X axis between the minimum and maximum values of data. The average value of the data within a bin is indicated with a red square symbol, and the green vertical solid line denotes its standard error bar. The locations of the red squares specify the right side of each bin on the X axis, thus the maximum value of the data along the X axis is the location of the last square. A weighted linear fit to the average values of the bins is shown with a red dashed line in each panel of Figure 5. The weight of each bin is determined by the number of the data in the bin. The method of binning and linear fits has been used by \citet{2009ApJ...702.1048W,2012ApJ...745..163K, 2012ApJ...752..109L}.

In the left column of Figure 5, the intensity--diameter (Figure 5a), lifetime--diameter (Figure 5c), and lifetime--intensity (Figure 5e) exhibit an increasing trend. However, the velocities present a weak inverse relationship to the intensity ratios (Figure 5g). The results demonstrate that larger CUDs tend to be brighter, live longer and move slower. The trends of PUDs are similar with CUDs, which can be found from the panels in the right column of Figure 5. The weighted correlation coefficients of the bins are listed in Table 4.

\citet{2010A&A...510A..12B} concluded from realistic radiative MHD simulations that the area--lifetime and brightness--lifetime relationships are positively correlated. They stated the reason is that stronger and more extended convective up-flows are maintained longer and create larger and brighter UDs. Similar trends were found by \citet{2002A&A...388.1048T,2009A&A...504..575S,2012ApJ...745..163K}. \citet{2012ApJ...745..163K} studied the statistical properties of UDs using high-resolution observations recorded by the \textit{New Solar Telescope} at the Big Bear Solar Observatory and three-dimensional MHD simulations of sunspots. They concluded that the UD velocities are inversely related to their lifetimes. Our finding as regards of the CUD and PUD trends also supports their conclusions.

\begin{figure}
        \centering
        \includegraphics[width=6cm]{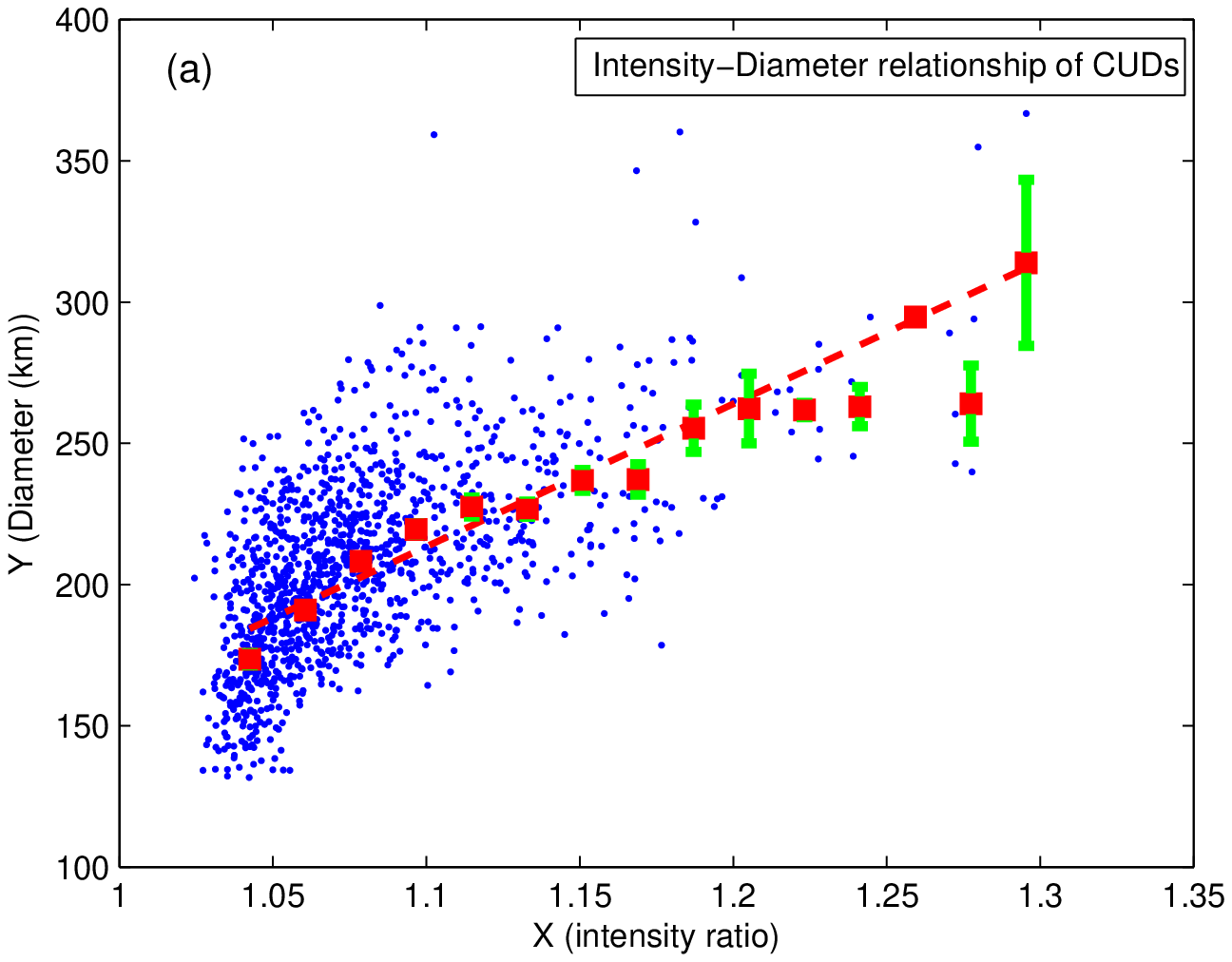}
        \includegraphics[width=6cm]{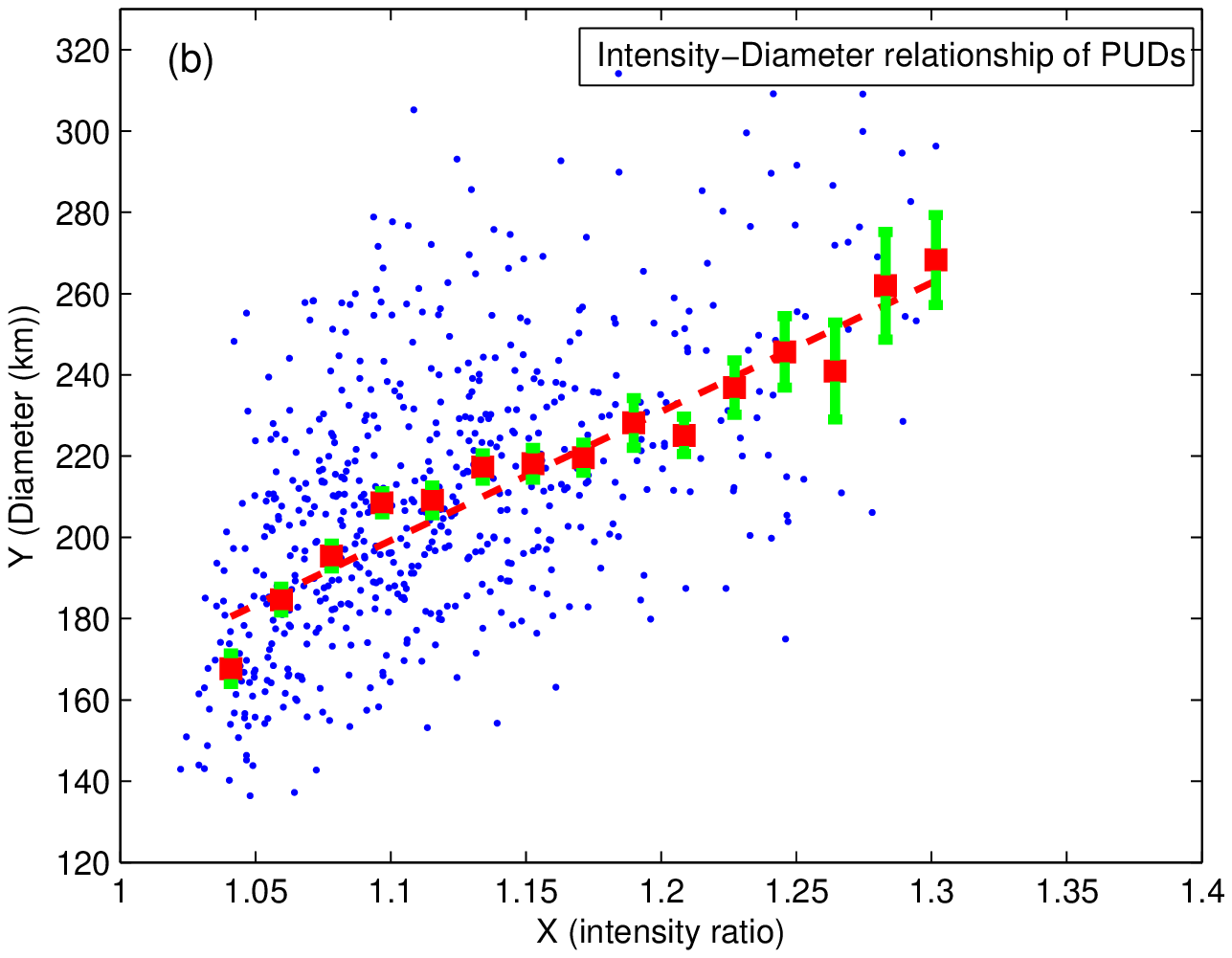}
        \includegraphics[width=6cm]{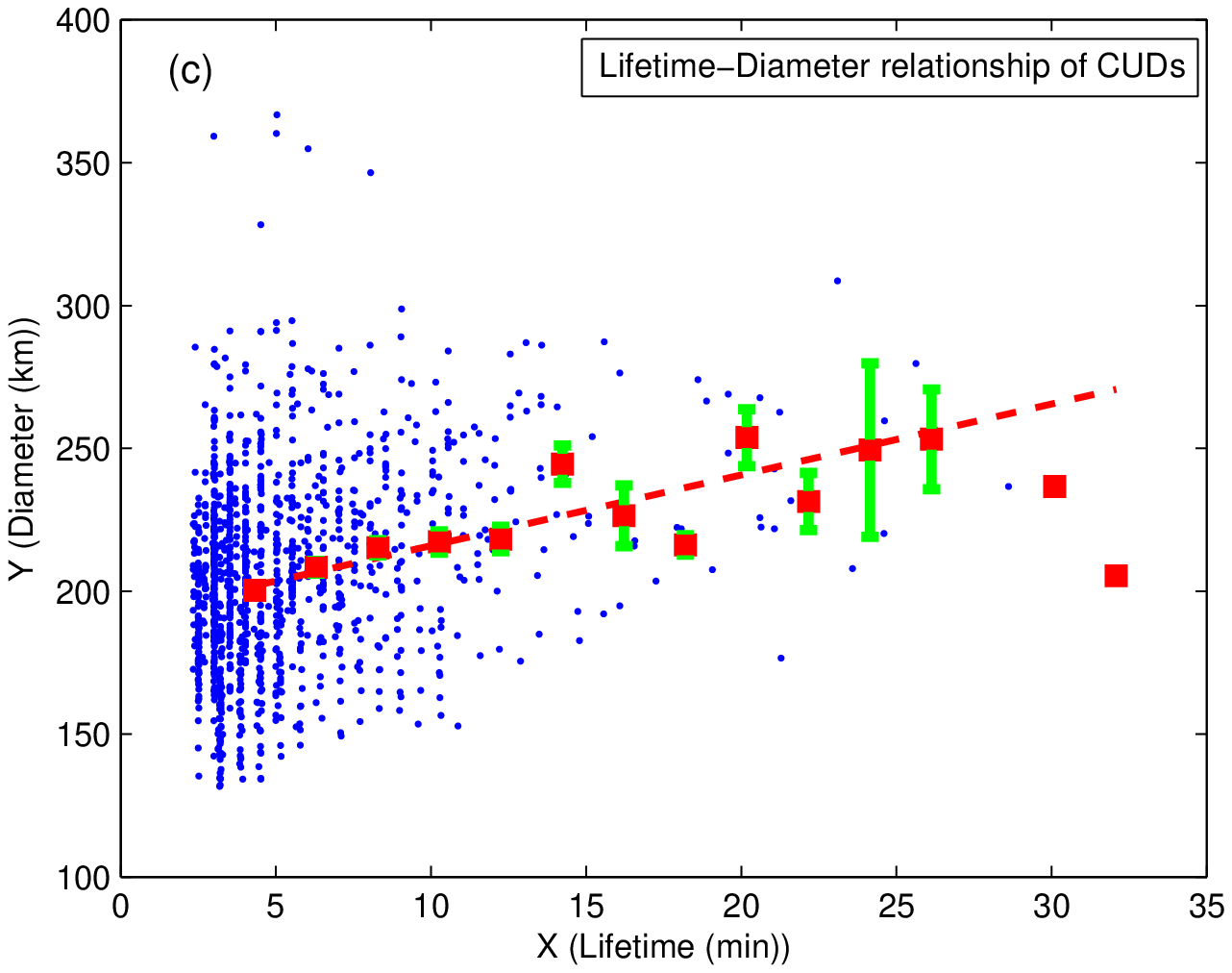}
        \includegraphics[width=6cm]{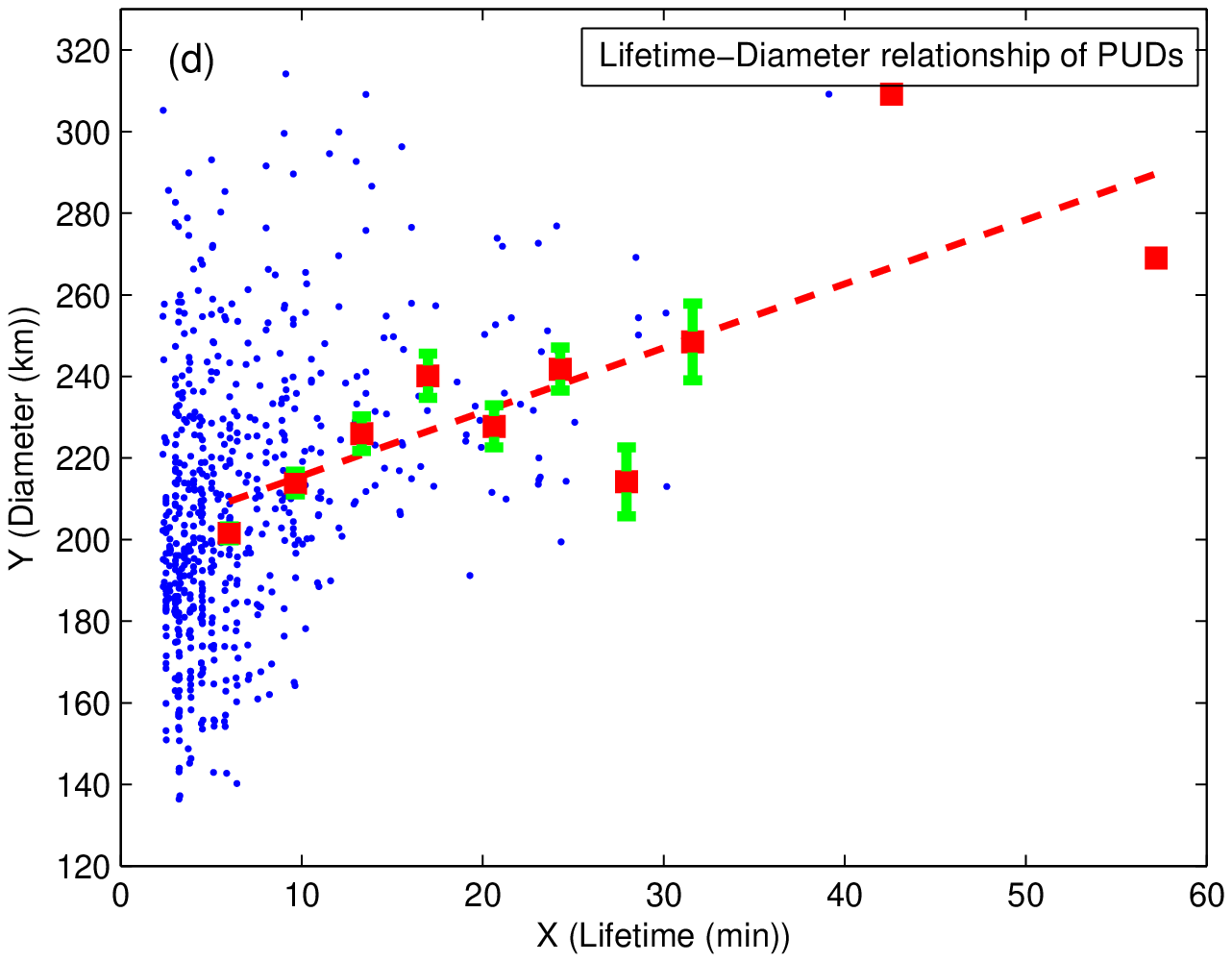}
        \includegraphics[width=6cm]{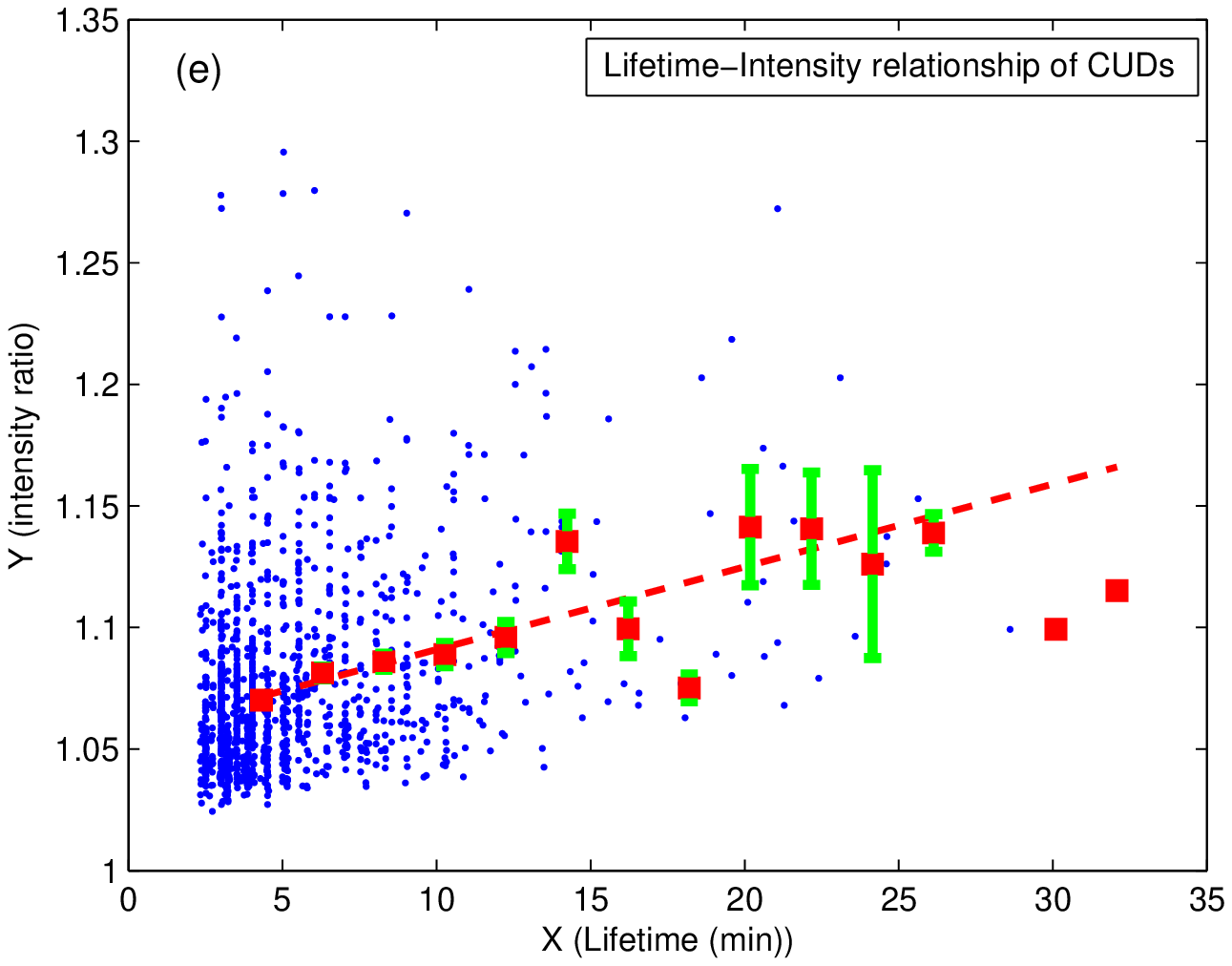}
        \includegraphics[width=6cm]{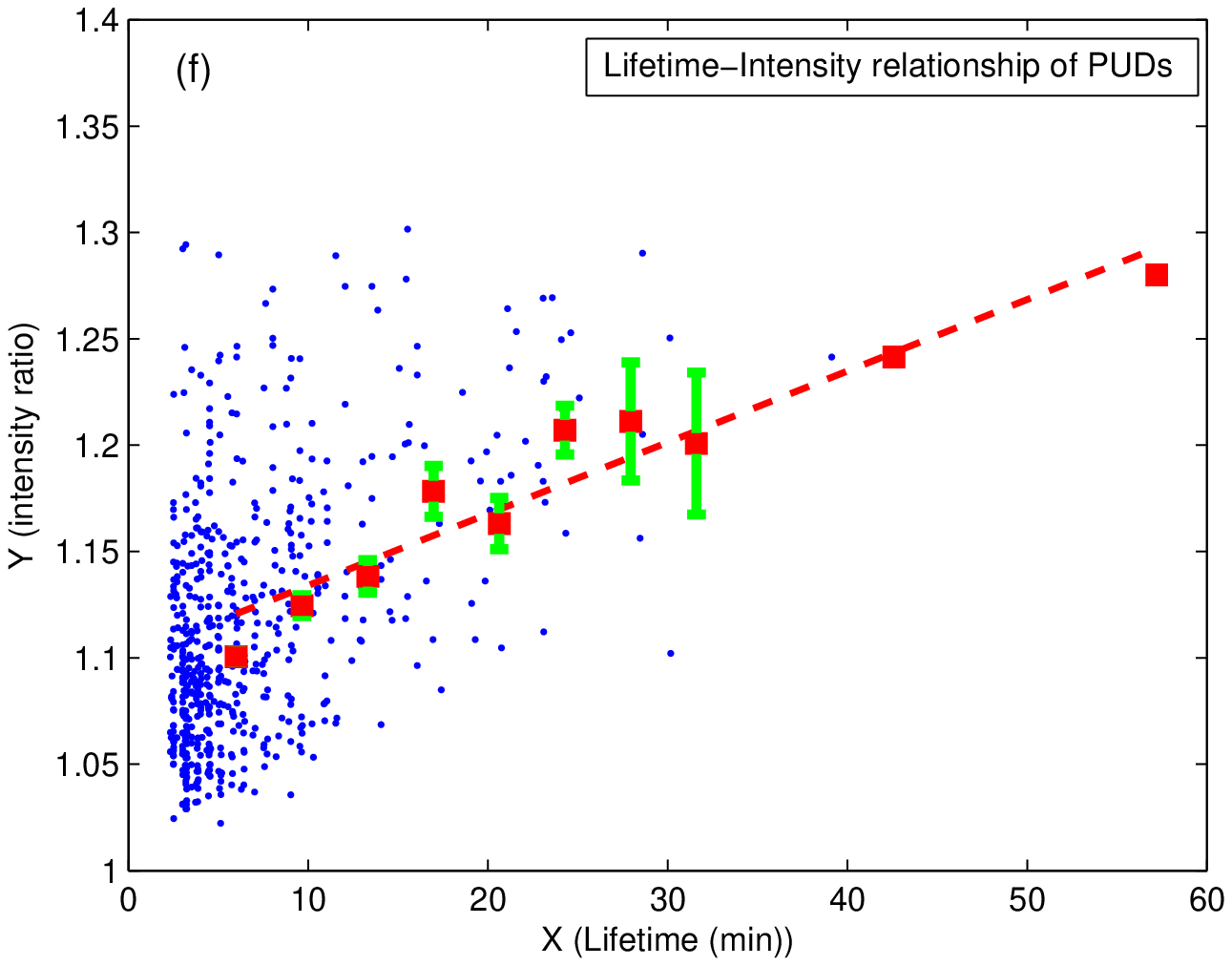}
        \includegraphics[width=6cm]{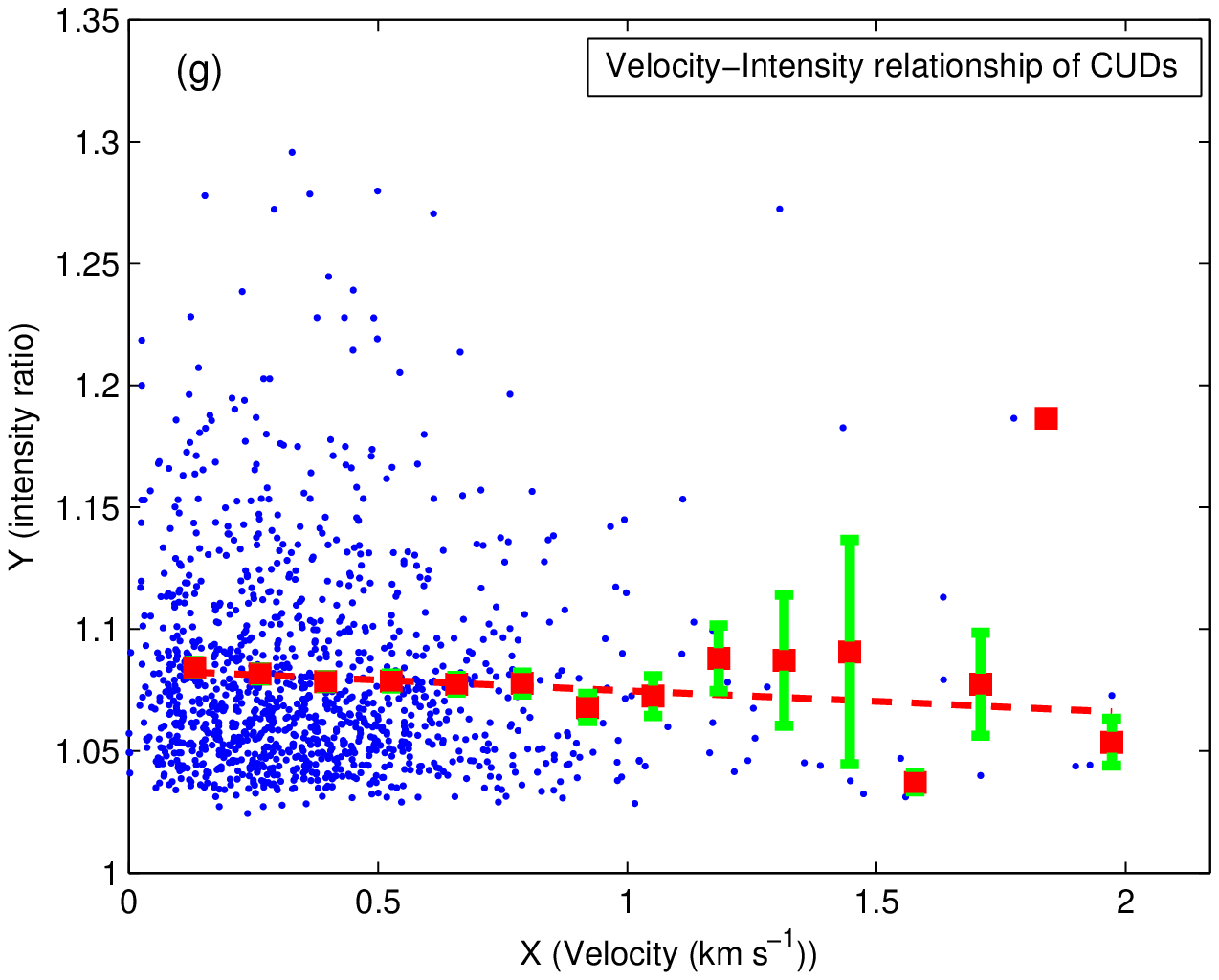}
        \includegraphics[width=6cm]{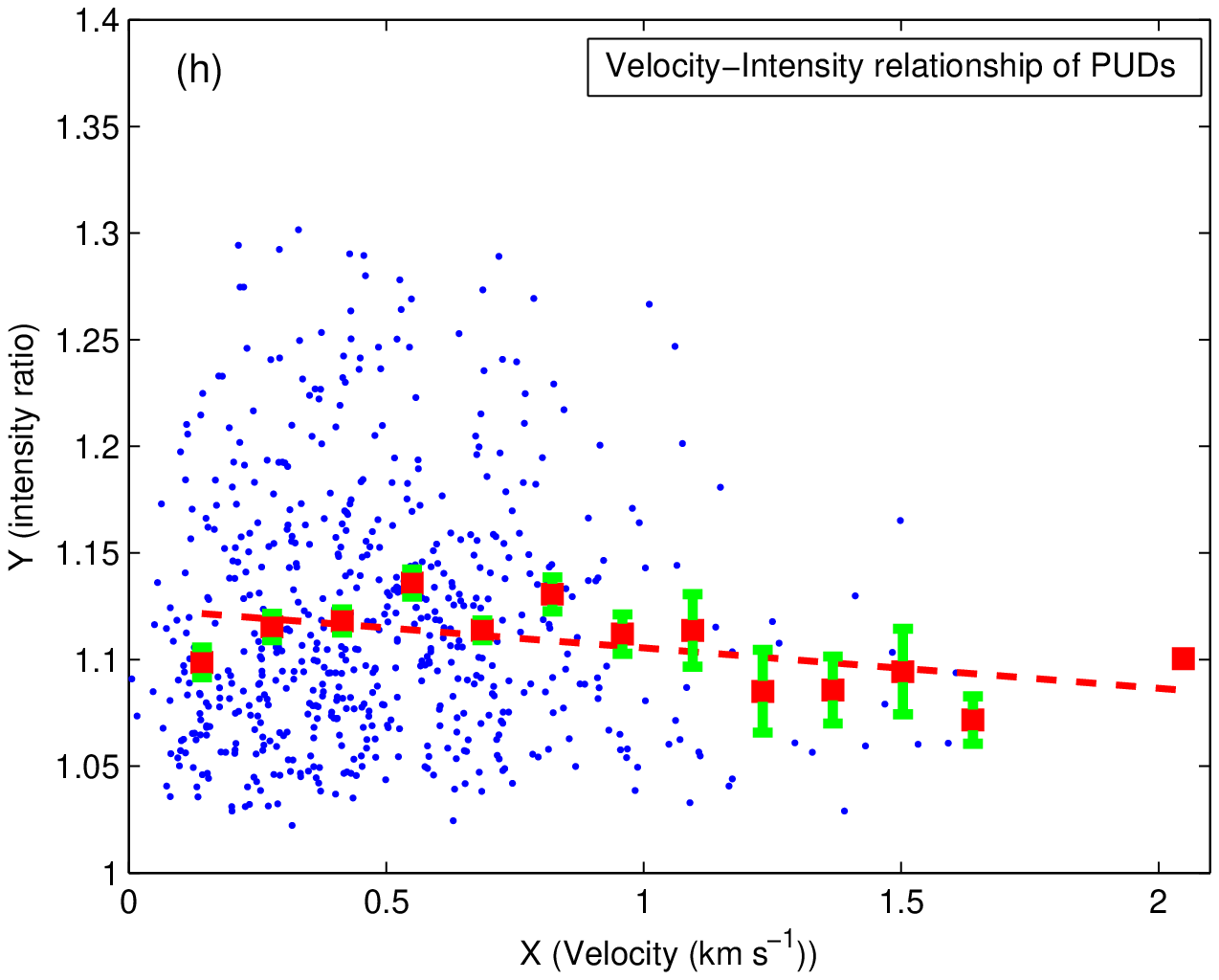}
        \caption{Scatter plots of all UDs obtained from the four ARs. The left column illustrates the CUD relationships and the right column the PUD ones. (a--b) Corresponds to the intensity--diameter relationships, (c--d) to the lifetime--diameter, (e--f) to the lifetime--intensity, and (g--h) to the velocity--intensity ones. The red square symbols indicate the average values of bins, and the green vertical solid lines denote their standard error bar. The red dashed lines are the weighted linear fit of the bins.}
        \label{fig5}
\end{figure}

\begin{table}
\caption{Weighted correlation coefficients of UD properties}
\begin{tabular}{ccccc}
\hline
    & Intensity--Diameter	&Lifetime--Diameter		&Lifetime--Intensity		&Velocity--Intensity\\
\hline
CUD		& 0.95	    &0.90      &0.89			&-0.40	\\
\hline
PUD	    & 0.95	    &0.92	   &0.97			&-0.03	\\

\hline
\label{tbl5}
\end{tabular}
\end{table}

\subsection{Relationships Between UD Properties and Umbral Magnetic Field Strengths}

In order to analyze relationships between UD properties and umbral magnetic field strengths, we used the radial component of the vector magnetic field, $B_r$ for the six sunspots taken with the \textit{Helioseismic and Magnetic Imager} on-board the \textit{Solar Dynamics Observatory} \citep[SDO/HMI,][]{2012SoPh..275..229S}.  Because the HMI image pixel-scale is 0.5" and UD diameters are approximately 0.3" \citep{2006ApJ...641L..73S,2012ApJ...745..163K,2015SoPh..290.1119F, 2015ApJS..219...17Y}, we failed to obtain an accurate pixel-by-pixel magnetic field feature for each UD. So we used the average magnetic field strength of the peripheral and central regions of each umbra. To improve the sensitivity of the average field strength, the $B_r$ maps were selected during the observed time interval of each data set and then averaged. Figure 6 shows two $B_r$ maps that were from NOAA 11801 and 12158. The maps were first extended to the scale of TiO images using a nearest-neighbor interpolation method and aligned to the corresponding corrected TiO image (\textit{i.e.,} the image shown in Figure 3). In this figure, the periphery and center boundaries are superposed on the maps with yellow and red curves. The average magnetic field strength of the peripheral and central regions are listed in Table 5.

Figure 7 shows the relationships between CUD properties (diameter, intensity, lifetime, and velocity) and umbral magnetic field strengths. In each panel of Figure 7, the six red square symbols indicate the property values of six CUD data sets obtained with six sunspots, and those corresponding green vertical solid lines denote their standard error bars; the red dashed line represents the best linear fit of the red squares. These correlation coefficients are -0.87, -0.91, -0.47, and 0.79 for the four fits in Figure 7a--d.

From the fitted lines in Figure 7 we find that diameters, intensities and lifetimes of CUDs have a decreasing trend with increasing magnetic field strength (see plots in Figure 7a--c), while velocities increase (see Figure 7d). The results demonstrate that CUDs are larger and brighter, and their lifetimes longer, however, their motions slower in a weaker umbral magnetic field environment than a stronger one. To PUDs, the three means of the properties obtained with three sunspots are so few that we did not obtain a significant result.

\begin{figure}
        \centering
        \includegraphics[height=4.5cm]{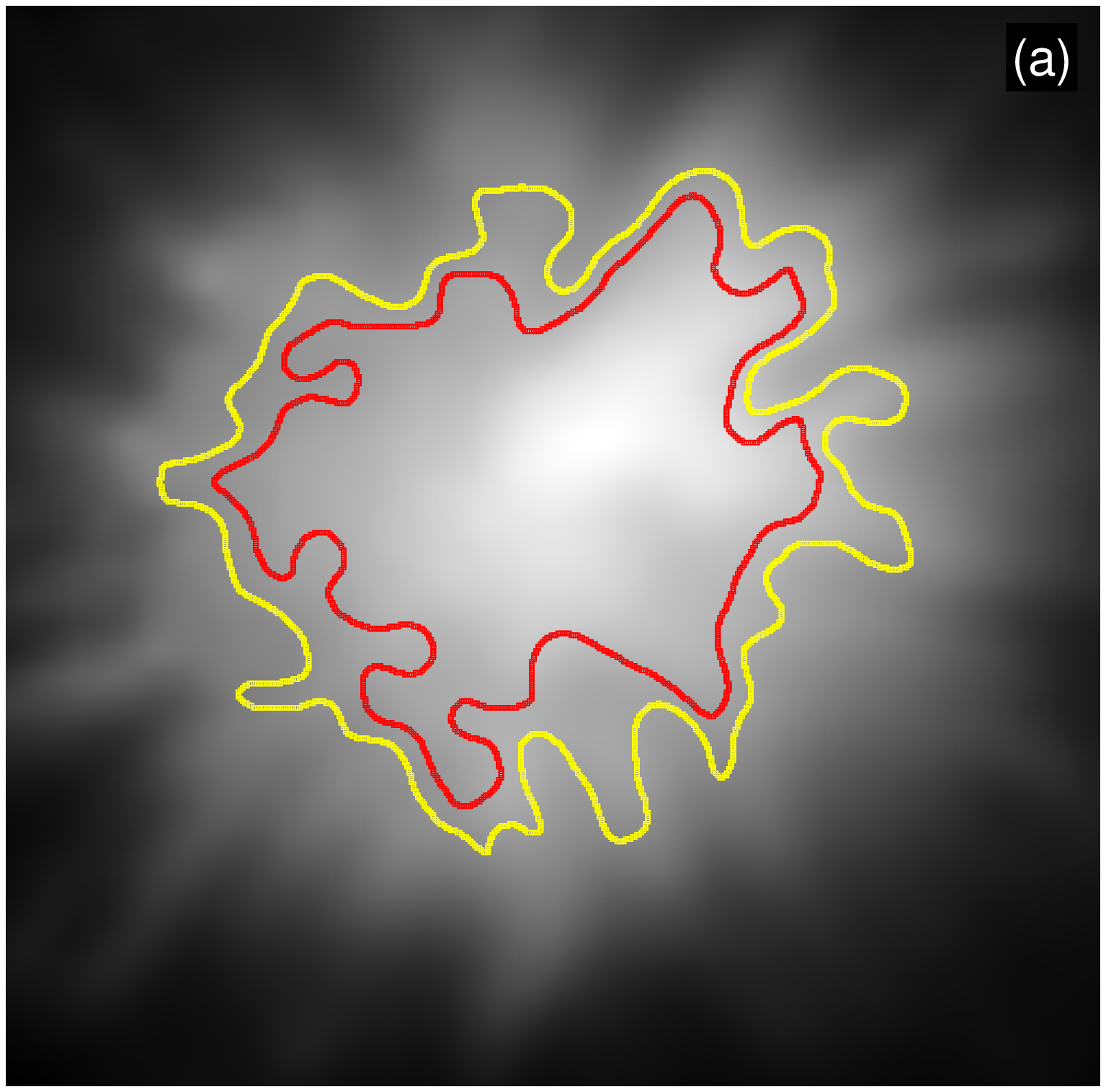}
        \includegraphics[height=4.5cm]{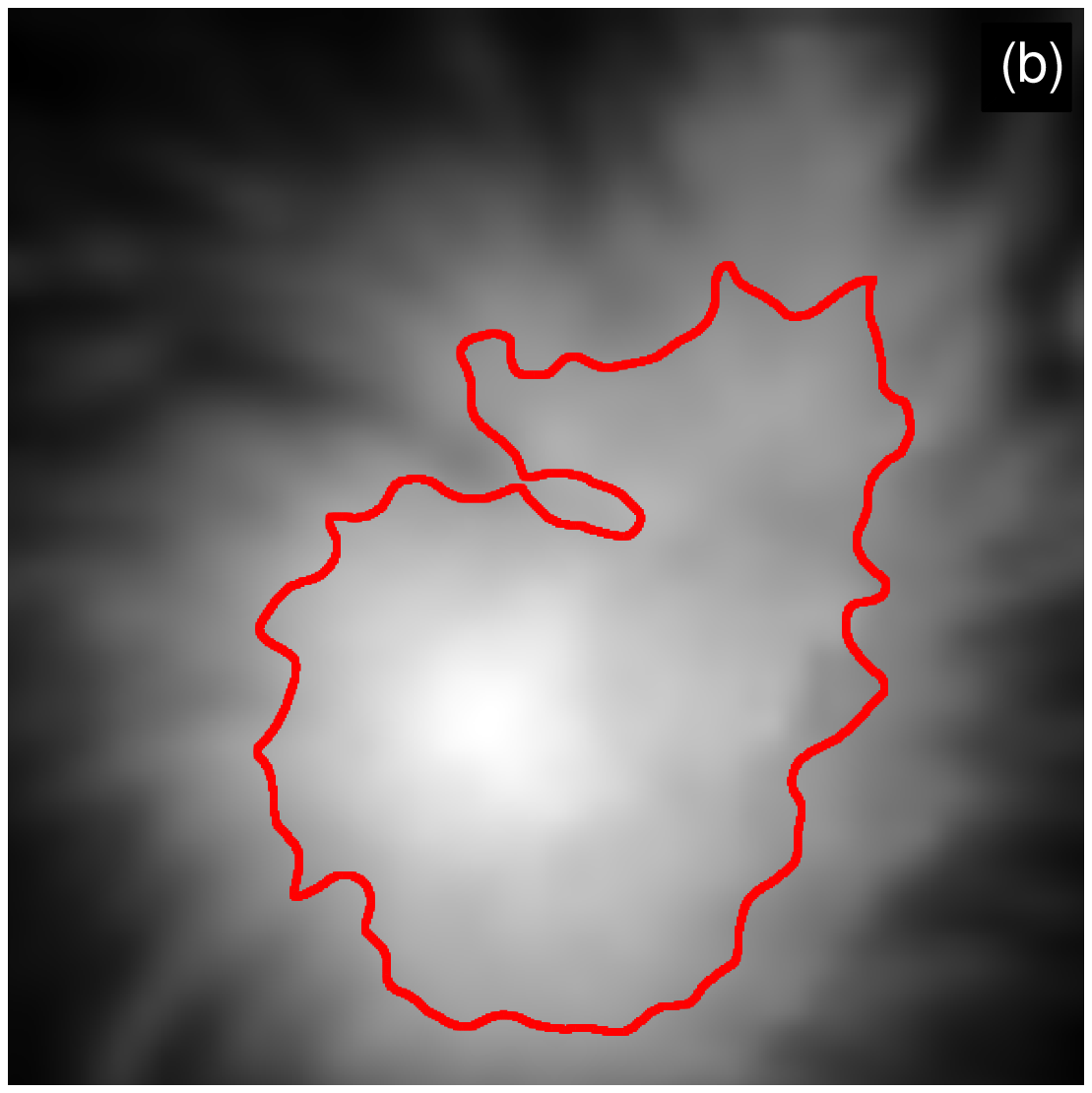}
        \caption{The radial component of the vector magnetic field $B_r$ superposed on the periphery (umbra--penumbra) and center boundaries with yellow and red curves: (a) $B_r$ map of NOAA 11801 obtained with the SDO/HMI on 1 August 2013 at 03:48:00 UT. (b) $B_r$ map of NOAA 12158 obtained on 13 September 2014 at 03:00:00 UT.}
        \label{fig6}
\end{figure}

\begin{table}
\caption{Mean magnetic field strengths (G) of each umbra in the central (surrounding CUDs) and peripheral (surrounding PUDs) regions.}
\begin{tabular}{ccccccc}
\hline
\multicolumn{7}{c}{CUDs}\\
\hline
AR NOAA     & 11598     &11801      &\multicolumn{3}{c}{12158}      &12178   \\
\hline
Spot         &           &           & A     &B      &C              &  \\
\hline
Mean Field Strength &2334 	    & 2009      & 1566&2079&1741&2212 \\
\hline
\multicolumn{7}{c}{PUDs}\\
\hline
Mean Field Strength &1888 	    & 1582      & &&&1688\\

\hline
\label{tbl6}
 \end{tabular}
\end{table}

\begin{figure}
        \centering
        \includegraphics[width=6cm]{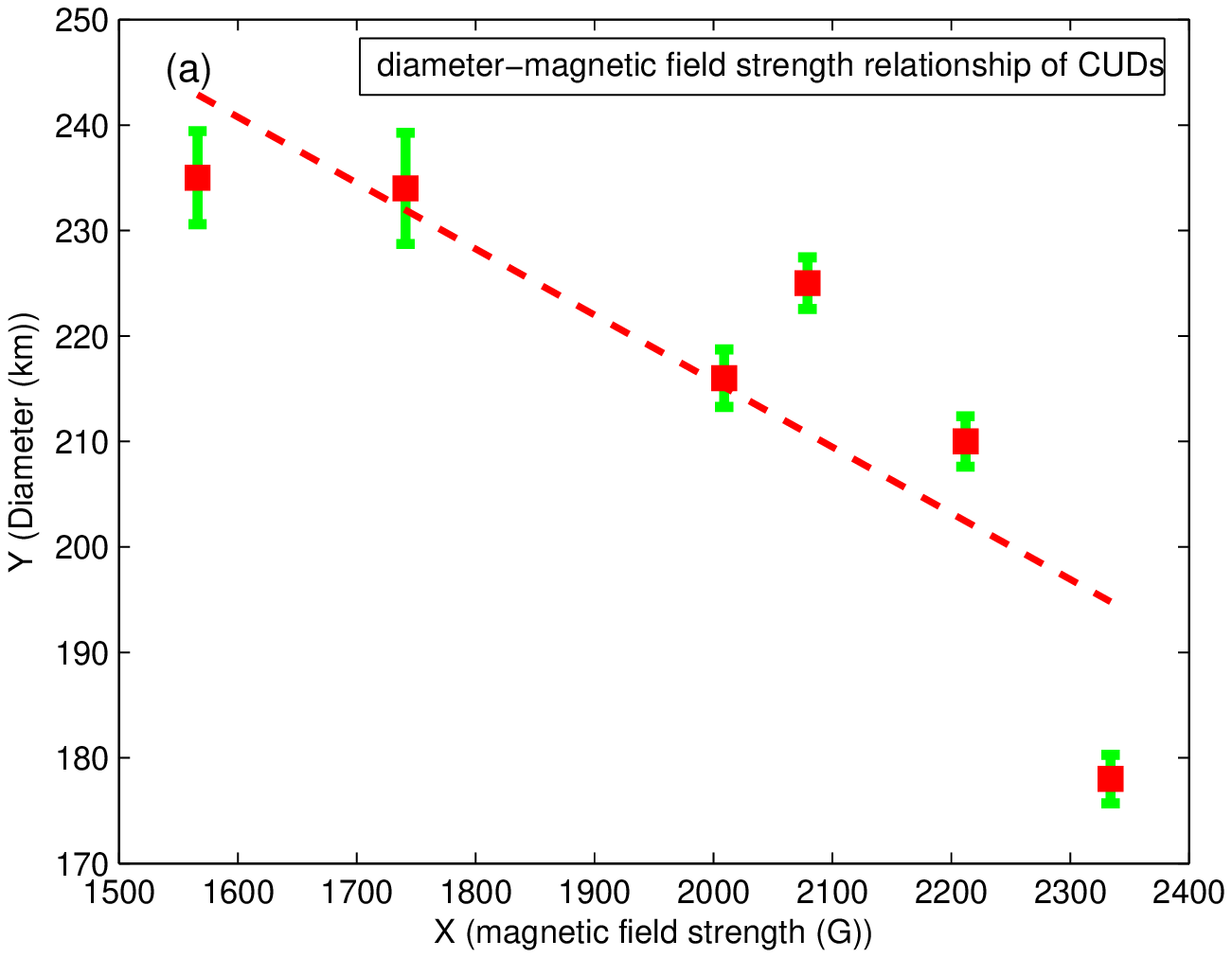}
        \includegraphics[width=6cm]{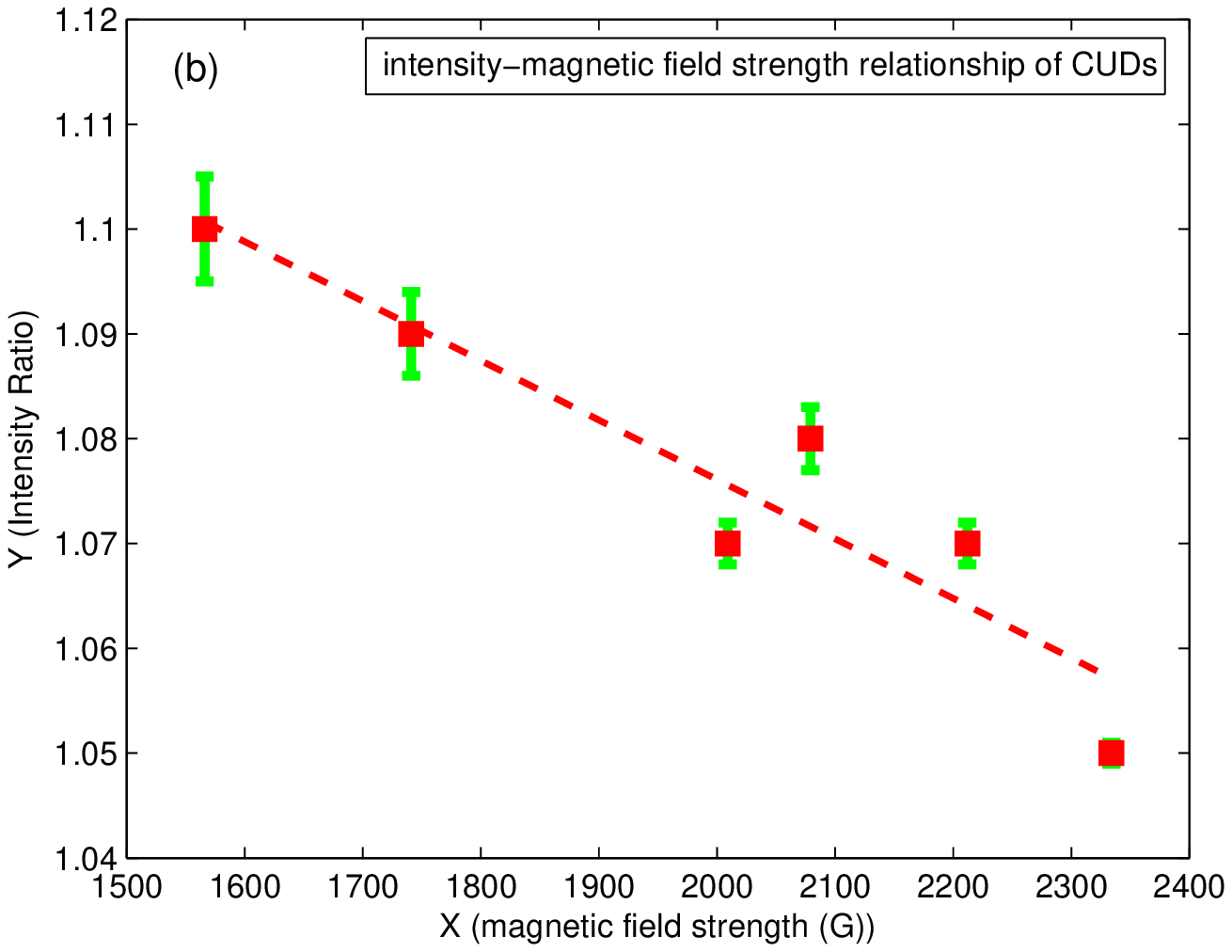}
        \includegraphics[width=6cm]{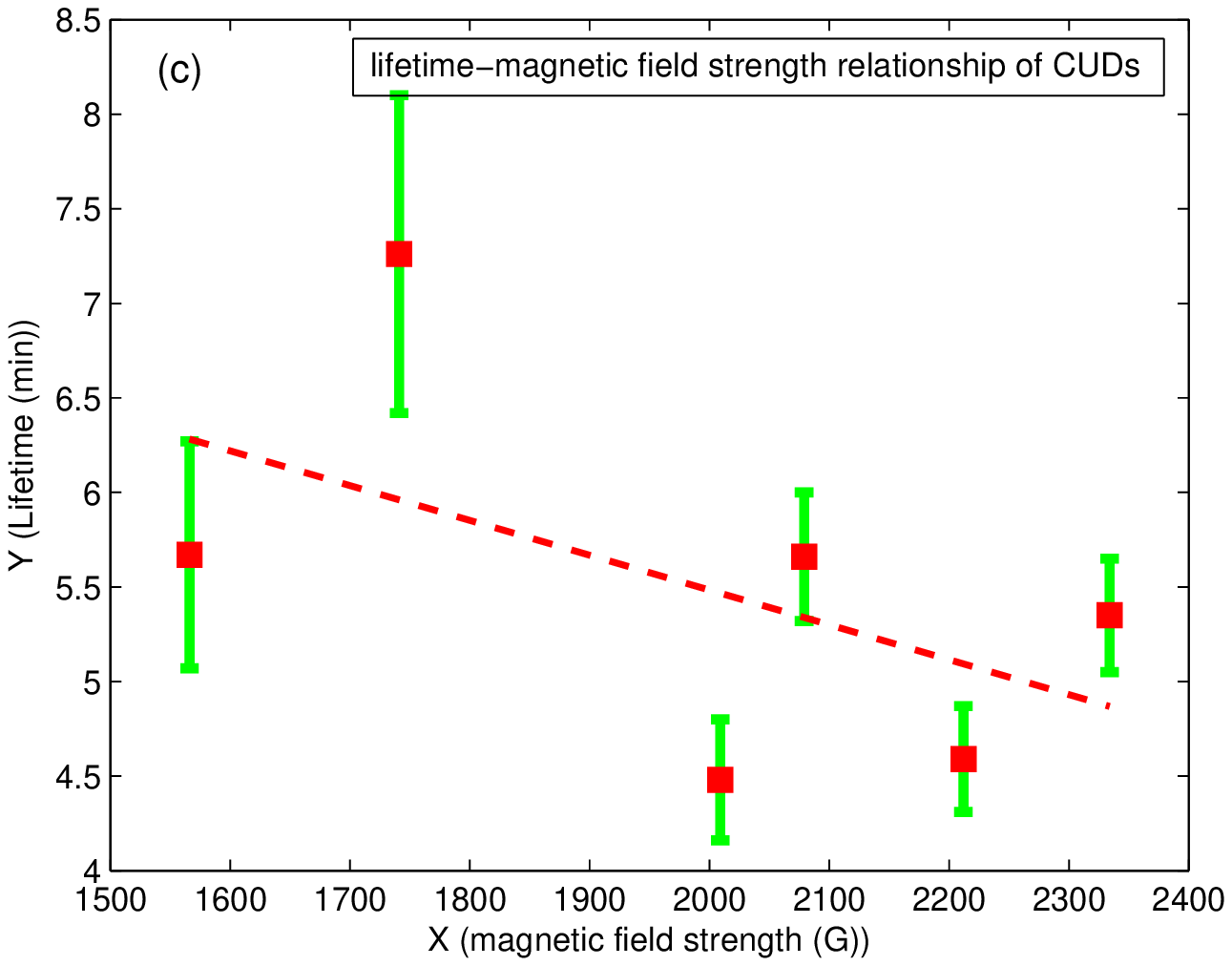}
        \includegraphics[width=6cm]{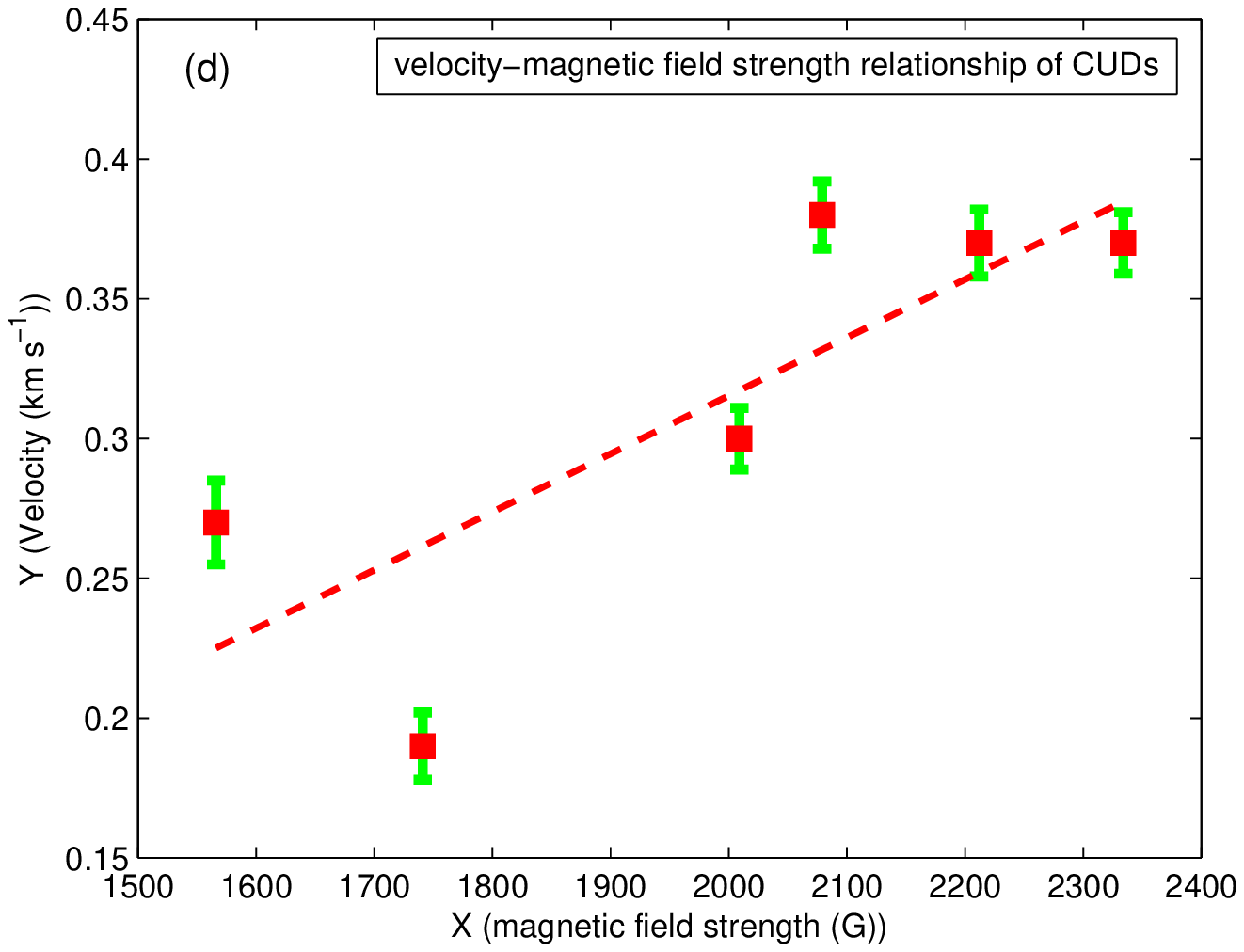}
        \caption{Relationships between CUD properties and umbral magnetic field strengths: (a) diameter--magnetic field strength plot, (b) intensity--magnetic field strength plot, (c) lifetime--magnetic field strength plot, and (d) velocity--magnetic field strength plot. The red square symbols indicate the means of CUD properties in each sunspot, and the corresponding green vertical solid lines denote their standard error bars; the red dashed lines represent the best linear fit to the property values.}
        \label{fig7}
\end{figure}

\section{Conclusions}
We selected high-resolution TiO image sequences of four ARs taken under the best seeing conditions with the NVST to investigate UD properties and analyze their relationships. Subsequently, umbral magnetic field strength relationships to UD properties were analyzed using the radial component of the vector magnetic field obtained from SDO/HMI.

We found that diameters and lifetimes of CUDs hold an increasing trend with the brightness, but velocities do not. The trends of PUDs are similar to those of CUDs. Moreover, UD properties depend on their corresponding magnetic field environment. A CUD diameter is larger, its brightness higher, its lifetime longer, and its velocity slower in a weak umbral magnetic field environment than in a strong one.

\begin{acks}
The authors are grateful to the referee and the editors for enlightening comments and suggestions, which improved the quality of our work. The authors thank the NVST team for their high-resolution observations. The HMI data used here are courtesy of NASA/SDO and the HMI science teams. This work is supported by the National Natural Science Foundation of China (Numbers: U1231205, U1531132, 11573012, 11463003, 11303011, 11263004, 11163004) and the Open Research Program of Key Laboratory of Solar Activity of the Chinese Academy of Sciences (Numbers: KLSA201414, KLSA201505). This work is also supported by the Opening Project of the Key Laboratory of Astronomical Optics and Technology, Nanjing Institute of Astronomical Optics and Technology, Chinese Academy of Sciences (Number: CAS-KLAOT-KF201306).
\end{acks}

\end{article}
\end{document}